\documentclass{emulateapj}
\usepackage{textcomp}
\usepackage{amsmath}
\usepackage{graphicx}
\usepackage[T1]{fontenc}

\def\beq{\begin{eqnarray}}
\def\eeq{\end{eqnarray}}

\begin{document}

\title{Fast radio bursts and their gamma-ray or radio afterglows as Kerr-Newman black hole binaries}

\author{Tong Liu,\altaffilmark{1,5} Gustavo E. Romero,\altaffilmark{2,3} Mo-Lin Liu,\altaffilmark{4} and Ang Li\altaffilmark{1,5}}

\altaffiltext{1}{Department of Astronomy, Xiamen University, Xiamen, Fujian 361005, China; tongliu@xmu.edu.cn, liang@xmu.edu.cn}
\altaffiltext{2}{Instituto Argentino de Radioastronom\'{i}a (IAR, CCT La Plata, CONICET), C.C.5, 1894 Villa Elisa, Buenos Aires, Argentina; romero@iar.unlp.edu.ar}
\altaffiltext{3}{Facultad de Ciencias Astron\'{o}micas y Geof\'{i}sicas, Universidad Nacional de La Plata, Paseo del Bosque s/n, 1900 La Plata, Argentina}
\altaffiltext{4}{College of Physics and Electronic Engineering, Xinyang Normal University, Xinyang, Henan 464000, China; mlliu@xynu.edu.cn}
\altaffiltext{5}{Department of Physics and Astronomy, University of Nevada, Las Vegas, NV 89154, USA}

\begin{abstract}
Fast radio bursts (FRBs) are radio transients lasting only about a few milliseconds. They seem to occur at cosmological distances. We propose that these events can be originated in the collapse of the magnetosphere of Kerr-Newman black holes (KNBHs). We show that the closed orbits of charged particles in the magnetosphere of these objects are unstable. After examining the dependencies on the specific charge of the particle and the spin and charge of the KNBH, we conclude that the resulting timescale and radiation mechanism fit well with the extant observations of FRBs. Furthermore, we argue that the merger of a KNBH binary is one of the plausible central engines for potential gamma-ray or radio afterglow following a certain FRBs, and can also account for gravitational wave (GW) events like GW 150914. Our model leads to predictions that can be tested by combined multi-wavelength electromagnetic and GW observations.
\end{abstract}

\keywords {black hole physics - gamma-ray bursts: general - gravitational waves - binaries: general}

\section{Introduction}

Fast radio bursts (FRBs) are transient astrophysical sources with radio pulses lasting only about a few milliseconds and a total energy release of about $10^{38}-10^{40}$ ergs. They are observed at high Galactic latitudes, and have anomalously high dispersion measure values \citep[e.g.,][]{Lorimer2007,Thornton2013,Katz2016}. So far, no electromagnetic counterpart has been detected in other frequency bands.

Several models have been recently introduced in the literature to explain the progenitors of FRBs. These models include magnetar flares \citep{Popov2010,Popov2013,Totani2013,Kulkarni2014,Lyubarsky2014}, annihilating mini black holes (BHs) \citep{Keane2012}, mergers of binary white dwarfs \citep{Kashiyama2013}, delayed collapse of supermassive neutron stars (NSs) to BHs \citep{Falcke2014}, flaring stars \citep{Loeb2014}, superconducting cosmic strings \citep{Yu2014}, relevant short gamma-ray bursts (GRBs) \citep{Zhang2014}, collisions between NSs and asteroids/comets \citep{Geng2015}, soft gamma repeaters \citep{Katz2015}, BH batteries \citep{Mingarelli2015}, quark nova \citep{Shand2016}, coherent Bremsstrahlung in strong plasma turbulence \citep{Romero2016}, and young supernova (SN) remnant pulsars \citep{Connor2016,Cordes2016}. Lately, FRB 140514 was found to be $21\pm7\%$ (3$\sigma$) circularly polarized on the leading edge with a 1$\sigma$ upper limit on linear polarization $<10\%$ \citep{Petroff2015}. This provides important constraints on the progenitors. In addition, FRBs may be used as a viable probe to constrain cosmography \citep[e.g.,][]{Gao2014,Zhou2014}. All in all, FRBs are among the most mysterious sources known in current astronomy.

In this paper we propose that FRBs can arise when a Kerr-Newman BH (KNBH) suddenly discharges. The process destroys the source of the magnetic field associated with the ergospheric motion of the electric field lines. The field then recombines at the speed of light coherently exciting the ambient plasma and producing a radio pulse. If the KNBH is part of a binary system, the instability is triggered by the tidal interactions in the pre-merging phase. This results in a FRB precursor of the gravitational wave (GW) burst.

The Kerr-Newman (KN) metric has been widely studied after Newman and Janis found the axisymmetric solution of Einstein's field equation for a spinning charged BH \citep{Newman1965}. In astrophysics, it is generally believed that a KNBH or a Reissner-Nordstr\"{o}m BH (RNBH) could not exist for a long time in a plasma environment because of the charge accretion would neutralize the BH on short timescales \citep{Ruffini1973}. However, the charge distribution in the magnetosphere can be time stationary when the rotation of the plasma balances the electrostatic attraction of the BH \citep{Punsly1998}. Once the mechanical equilibrium is broken because of the magnetosphere instability, then the electromagnetic energy can be released from the KNBH.

KNBHs have had only limited applications in astrophysics so far:  they were invoked to explain some unidentified, low-latitude, gamma-ray sources early observed by EGRET \citep[e.g.,][]{Punsly2000,Eiroa2002,Torres2001,Torres2003} and gravitational lensing effects \citep[e.g.,][]{Kraniotis2014}.

In what follows we focus on the magnetosphere instability of a KNBH and its possible consequences related to FRBs and their potential afterglows. In Section 2, we describe the initial state of a KNBH, calculate the unstable orbits of a charged test particle surrounding a KNBH, plot the falling trajectories of a test particle, and estimate the corresponding discharge timescale. The radiation mechanism is discussed in Section 3. In Section 4, we briefly mention that the merger of a KNBH binary is one of the plausible central engines of FRBs and their possible afterglows. A short discussion and conclusions are presented in Section 5 and 6, respectively.

\section{Model}
\subsection{Unstable orbits of a test charged particle}

\begin{figure}
\centering
\includegraphics[angle=0,scale=0.52]{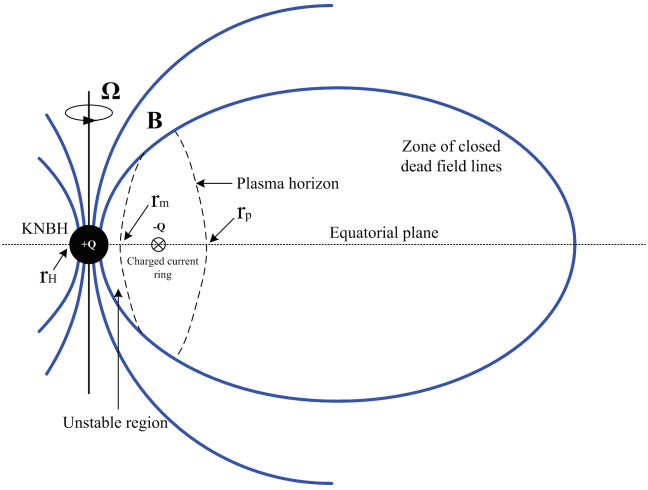}
\caption{Schematic diagram of the initial state of a KNBH.}
\label{fig1}
\end{figure}

For simplicty, we discuss the unstable orbits of a test charged particle in the magnetosphere. In the geometric unit system ($G=c=1$), the KN spacetime with mass $M$, angular momentum $J$, and electric charge $Q$ can be written in Boyer-Lindquist coordinates as \citep{Misner1973}
\beq
ds^2 &&= - \frac{\Delta}{\rho^2} \left[d t - a\sin^2\theta d \phi\right]^2\nonumber \\ &&+\frac{\sin^2 \theta}{\rho^2}\left[\left(r^2 + a^2\right)d \phi - a d t\right]^2+ \frac{\rho^2}{\Delta} d r^2 + \rho^2 d \theta^2,
\eeq
where
\beq
\Delta = r^2 - 2 M r + a^2 + Q^2,
\eeq
\beq
\rho^2 = r^2 + a^2 \cos^2\theta,
\eeq
and $a = J/M$ is angular momentum per unit mass. According to $\Delta=0$, the KNBH horizon can be defined as
\beq
r_{\rm H}=M+\sqrt{M^2-a^2-Q^2}.
\eeq
For a KNBH, the mass, spin and charge should satisfy the relation $M^2\geq a^2+Q^2$. Furthermore, the angular velocity of the horizon is:
\beq
\Omega_{\rm H}=\frac{a}{r_{\rm H}^2+a^2}.
\eeq
Here we just discuss the case of $M>a\gg Q>0$.

Following the notation of \citet{Misner1973}, the electromagnetic vector potential is
\beq
\emph{\textbf{A}}= (-\frac{Qr}{\rho^2},\  0,\  0, \frac{Qra \sin^2\theta}{\rho^2}),
\eeq
where the bold face means the vector. The electromagnetic vector potential $\emph{\textbf{A}}$ depends on the charge $Q$ and the specific angular momentum $a$ \citep{Hackmann2013}. The magnetic field is generated by the rotation of the charge distribution and the co-rotation of the charged BH electric field in the ergosphere.

The motions of the neutral test particles in the gravitational field or KN spacetime have been studied in some recent papers \citep[e.g.,][]{Liu2009,Liu2010,Liu2011,Pugliese2013}. Let a test particle of rest mass $m$ with charge $e$ be outside a KNBH and let us restrict ourselves to the case of orbits on the equatorial plane $\theta = \pi/2$. The contravariant components of the test particle's four-momentum (namely Carter's equations, see \citet{Carter1968}), $p^\alpha=dx^\alpha/d\lambda$, on the equatorial plane can be expressed as \citep{Misner1973}
\beq
p^\theta=0,
\eeq
\beq
r^2 p^r=\sqrt{R},
\eeq
\beq
r^2 p^\phi=-(aE-L_z)+\frac{a}{\Delta} P,
\eeq
\beq
r^2 p^t=-a(aE-L_z)+\frac{r^2+a^2}{\Delta}P,
\eeq
and the function of $R$ and $P$ on the equatorial plane are defined by
\beq
R=P^2-\Delta [m^2 r^2+(L_z-aE)^2],
\eeq
\beq
P=(r^2+a^2)E-a L_z-e Q r,
\eeq
where $L_z$ is axial component of angular momentum of the test particle. According to Eqs. (7-10), we can calculate the falling timescale and describe the infalling trajectories of the test particle on the equatorial plane.

\begin{figure}
\centering
\includegraphics[angle=0,scale=0.32]{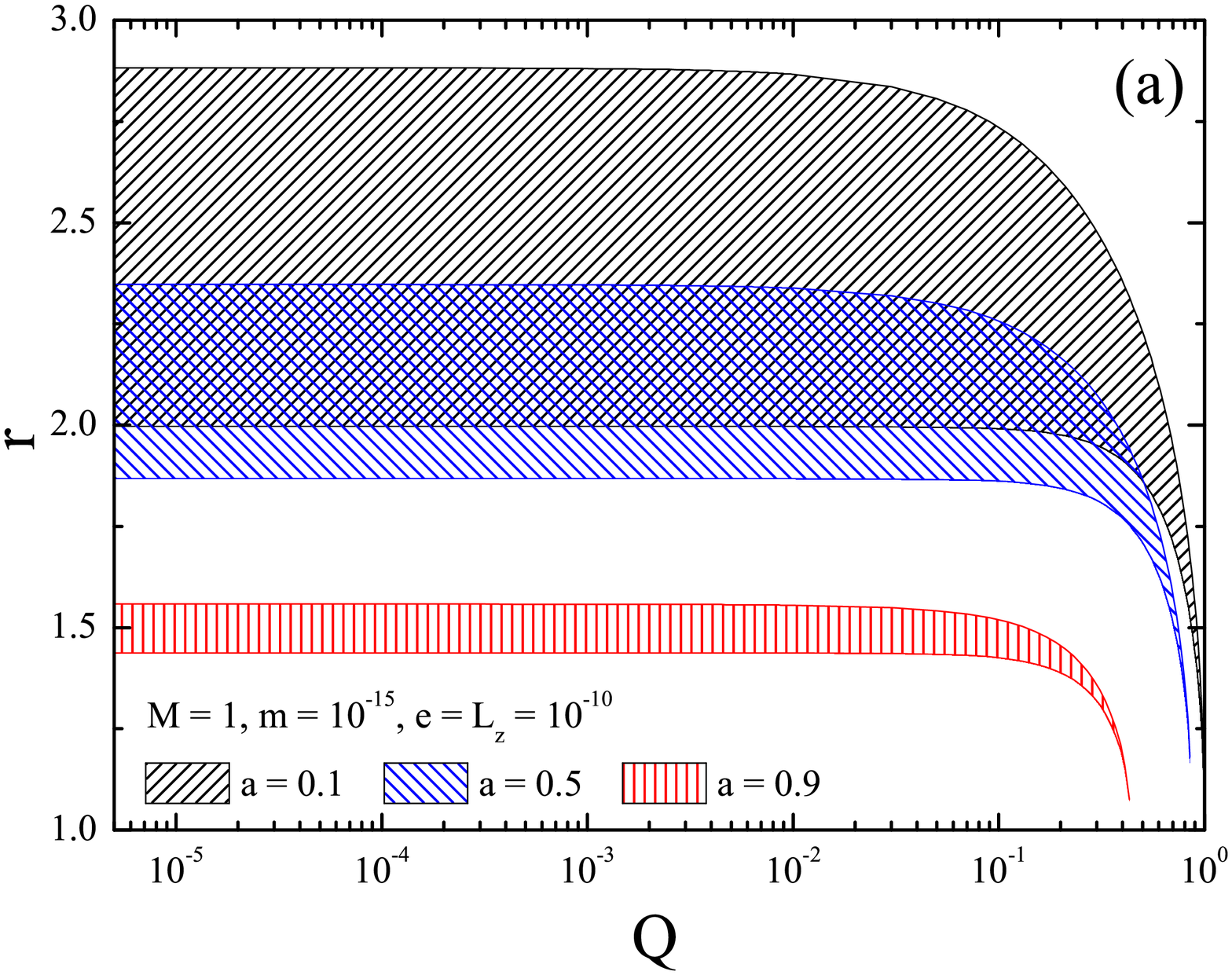}
\includegraphics[angle=0,scale=0.34]{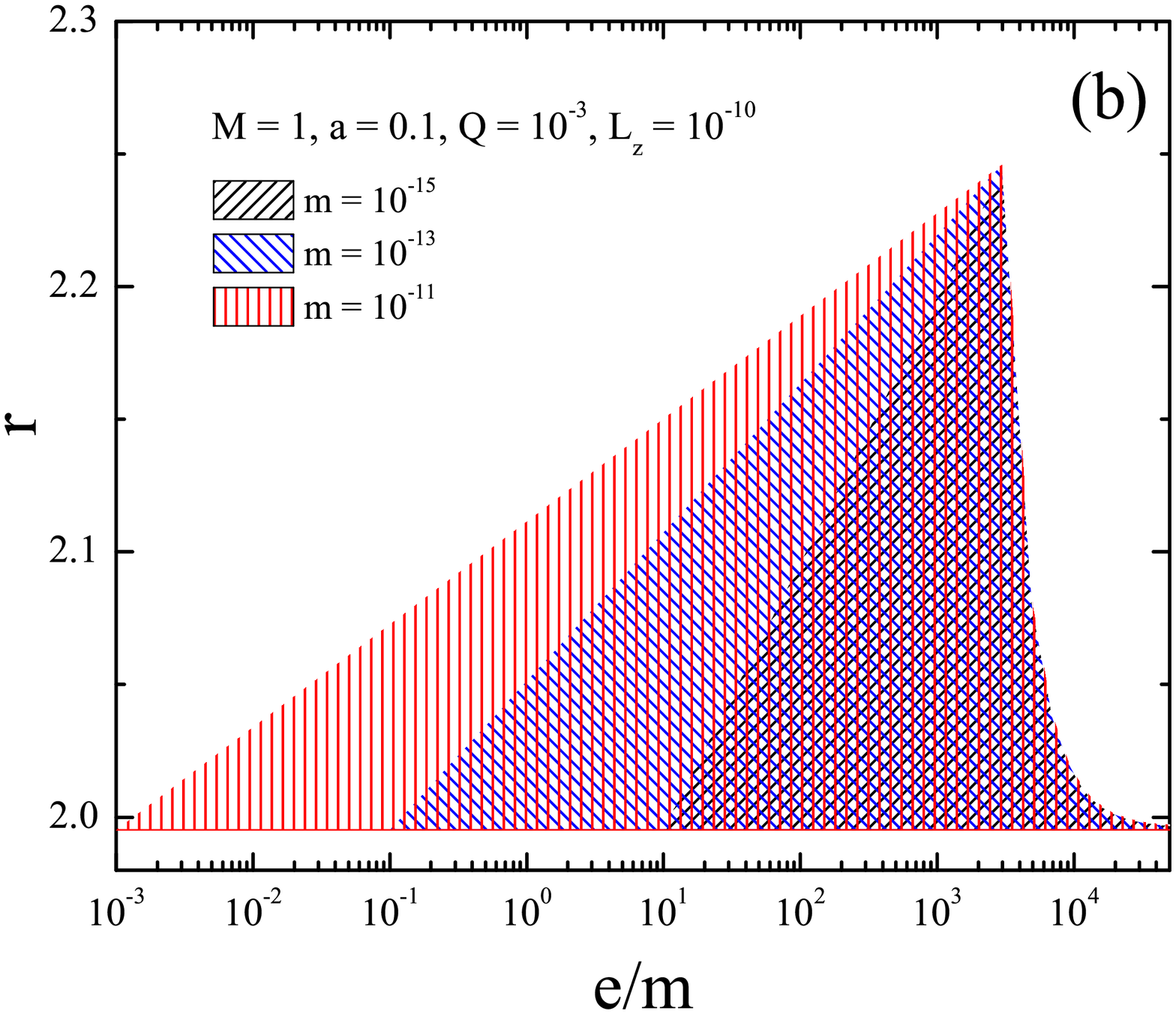}
\caption{Unstable region on the equatorial plane ($r_{\rm H}<r<r_{\rm m}$) around a KNBH ($M=1$) as functions of the KNBH spin and charge (panel a) with $m=10^{-15}$ and $e=L_z=10^{-10}$, and the specific charge of the particle (panel b) with $a=0.1,~Q=10^{-3}$, and $L_z=10^{-10}$.}
\label{fig2}
\end{figure}

From the equation of the radial momentum $p^r$, given by Eq. (8), the effective potential approach can then be adapted to study the dynamics of the particle. The radial motion is governed by the energy equation,
\beq
E = \frac{\beta}{\alpha} + \frac{\sqrt{\beta^2 - \alpha \gamma_0 + \alpha r^4 (p^r)^2}}{\alpha},
\eeq
where $\alpha$, $\beta$, $\gamma_0$ are functions of $r$ and of constants of motion written as follows:
\beq
\alpha = \left(r^2 + a^2\right)^2 - a^2\Delta > 0,
\eeq
\beq
\beta = \left(L_z a + e Q r\right)(r^2 + a^2) - L_z a \Delta,
\eeq
\beq
\gamma_0 = \left(L_z a + e Q r\right)^2 - L_z^2 \Delta- m^2 r^2 \Delta.
\eeq

Qualitative features of the radial motion can be derived from the effective potential $V(r)$, which is given by the minimum allowed value of $E$ at radial coordinate $r$,
\beq
V(r) = \frac{\beta}{\alpha} + \frac{\sqrt{\beta^2 - \alpha \gamma_0}}{\alpha}.
\eeq

The circular orbits can be deduced from the equation
\beq
\quad \frac{dV}{dr}=0,
\eeq
and the unstable orbit condition is given by
\beq
\frac{d^2 V}{d r^2} < 0.
\eeq

Here we define $r_{\rm m}$, which is satisfied with $dV/dr=d^2V/dr^2=0$, thus the unstable orbits on the equatorial plane are in the range between $r_{\rm m}$ and the KNBH horizon $r_{\rm H}$ as shown in Figure 1. The units of $r$ is $GM/c^2$ (or $r_{\rm g}/2$). $r_{\rm m}$ should be larger than the marginally stable circular orbit to ensure the test particle is out of the horizon. In the following descriptions, we use the normalized units until the BH mass $M_{\rm BH}$ is given in the units of $M_\odot$.

\subsection{Initial state of a KNBH}

The initial steady state configuration of a KNBH is shown in Figure 1. The bulk of the opposite charges of the magnetosphere forms an equatorial current ring, which exists in an area wrapped by a plasma horizon, corresponding to the radius with $r_{\rm m}<r<r_{\rm p}$ on the equatorial plane. The cause is that the quadrupole moment of the electric field dominates at radii larger than that of the ring, while the magnetic field is dipolar. At a large enough radius, the particles can exist in \emph{\textbf{E}}$\times$\emph{\textbf{B}} (\emph{\textbf{E}} and \emph{\textbf{B}} are the strengthes of electric and magnetic fields) drift trajectories and are not sucked into the KNBH.

From the plasma equilibrium condition, $r_{\rm p}$ should meet:
\beq
\left[ \frac{\Delta}{(r^2+a^2)^2-a^2\Delta \sin^2 \theta}\right]^{\frac{1}{2}}r^2 \rho \sin \theta|_{r= r_{\rm p}}= \frac{Q}{B},
\eeq
which is consistent with the results of RNBHs for $a=0$ \citep[e.g.,][]{Hanni1975,Damour1978,Karas1991}. Here $B$ is the modulus of the magnetic field. If we assume that the ring is located at $r=10M$, and $B=2aQ/r^3$ on the equatorial plane, thus $r_{\rm p}$ is about 23 $r_{\rm H}$ for $Q \ll a \sim M$. In addition, the closed dead field lines, shown in Figure 1, avoid the KNBH from spontaneous electric discharge. This point has been studied in detail in \citet{Punsly1998}.

\subsection{Results}

Figure 2 shows the unstable region on the equatorial plane ($r_{\rm H}<r<r_{\rm m}$) around a KNBH ($M=1$) as a function of the KNBH spin and charge (panel a) for $m=10^{-15}$ and $e=L_z=10^{-10}$, and the specific charge of the particle (panel b) for $a=0.1,~Q=10^{-3}$, and $L_z=10^{-10}$ (normalized units). From Figure 2 (a), we can see that for a test particle the size of the unstable regions decreases with the increase of the KNBH spin, and are almost independent of the KNBH charge up to its value is close to $\sqrt {M^2-a^2}$. Figure 2 (b) displays the constrain of the unstable regions on the characteristics of the test particle. The unstability conditions require the high-mass particles to have larger values of the charge. In such a case, the specific charge must be less than about $4 \times 10^4$ for the particles with different masses. According to Figure 2, the resulting unstable orbits are reasonably lying in the range of 1.5-3.

Since the detected FRBs have variability on millisecond timescales, which indicates that the emission region of FRBs is very compact, the BH mass can then be restricted within a few dozen times the solar mass. For a stellar-mass KNBH of $M_{\rm BH} \sim 20~M_\odot$, the unstable orbit for a charged particle is calculated to be about $10^7-10^8$ cm, and the unstable timescale can be estimated to be of $\sim$ 1 millisecond, which is the typical timescale of FRBs. Perhaps the charged particles distribute above or below the equatorial plane of KNBH, thus the unstable orbits may be larger than the orbits for the rest particle on the equatorial plane of KNBH, which lead to the falling timescale more in line with the FRB time.

By using Eqs. (8-9), the falling trajectories of a test particle can be plotted. Figure 3 shows the trajectories of a test particle ($m=10^{-15}$, and $e=L_z=10^{-10}$) on the equatorial plane falling into a KNBH ($M=1$, $a=0.1$, and $Q=10^{-3}$) from $r_0 =2.5$ to the inner boundary ($r= r_{\rm g}$) for $E=1.93 \times 10^{-11}$, $2 \times 10^{-11}$, and $2.4 \times 10^{-11}$. From Eq. (13), the minimum value of E is about $1.92 \times 10^{-11}$ at $r_0=2.5$ in this case. According to Eq. (10), the corresponding falling timescales can be calculated as 15.19, 13.28, and 11.26, respectively. For the BH mass $M_{\rm BH} \sim 20~M_\odot$, the falling timescale is about 1 ms, which coincides with the FRB timescale.

\begin{figure}
\centering
\includegraphics[angle=0,scale=0.4]{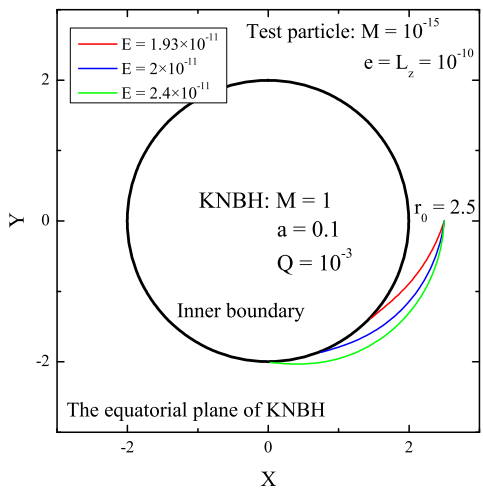}
\caption{The trajectories of a test particle ($m=10^{-15}$, and $e=L_z=10^{-10}$) on the equatorial plane falling into a KNBH ($M=1$, $a=0.1$, and $Q=10^{-3}$) from $r_0 =2.5$ to the inner boundary ($r= r_{\rm g}$) for $E=1.93 \times 10^{-11}$, $2 \times 10^{-11}$, and $2.4 \times 10^{-11}$.}
\label{fig3}
\end{figure}

\section{Radiation mechanism}

The electromagnetic structure of KNBHs is similar to that of NSs in pulsars. However, there are two major differences between them. First, BHs have no solid surfaces and consequently there is no thermal emission \citep{Punsly2000}. Second, for KNBHs, the rotation axis and magnetic axis are always aligned. KNBH, then, are non-pulsating sources. These features can be used to differentiate them from NSs.

\citet{Totani2013} suggested that binary NS mergers are a possible origin of FRBs, and the radiation mechanism is coherent radio emission, like in radio pulsars. \citet{Falcke2014} proposed the alternative scenario of a supermassive NS collapsing to a BH. In such a case, the entire magnetic field should in principle detach and reconnect outside the horizon. This results in large currents and intense radiation when the resulting strong magnetic shock wave moves at the speed of light through the remaining plasma. This very same mechanism should operate immediately after the discharge of a KNBH. For a magnetic field strength of $\sim 10^{12}-10^{13}$ G\, the expected energy-loss rate of KNBHs can meet the requirements of FRBs \citep{Falcke2014}. Also, as in the case of NSs, the radiation from KNBHs can bring the observed polarizations.

If the period of the KNBH is $P$, which is related to the BH mass and spin, i.e., $P\approx4\pi G M_{\rm BH}(1+\sqrt{1-a_*^2})/a_* c^3 $ for $a_*=a/M$ and $a\gg Q$ from Eq. (5), the size of its magnetosphere will be
\beq
R_{\rm mag}=c/\Omega \approx 4.8 \times 10^9 P ~\rm cm.
\eeq
For $P=0.01$ s ($a\approx$ 0.24 for BH mass $M_{\rm BH} \sim 20~M_\odot$), $R_{\rm mag}\sim 5 \times 10^6$ cm and the magnetic shock wave will collective excite the plasma in $\sim 0.5$ ms.

The curvature radiation power emitted per charge is
\beq
P_e=2 \gamma^4 e^2 c / 3 R_{\rm mag}^2,
\eeq
and the corresponding frequency is
\beq
\nu \approx 7 \gamma^3 [R/(10^{6} \; {\rm cm})]^{-1} ~\rm kHz.
\eeq
The bulk of the observed radio emission is then generated by particles with $\gamma\sim 1000$. This radiation is well above the relativistic plasma frequency $\nu_{\rm p}=\gamma^{-3/2}(4 \pi n_e e^2/ m_e)^{1/2}\sim 1$ GHz. The total power of the coherent pulse will be $P_{\rm tot}\sim (n_eV)^2 P_e$, where $V$ is the volume occupied by the plasma of density $n_e$ (for coherent curvature radiation see \citet{Ruderman1975} and \citet{Buschauer1976}). Typically, $P_{\rm tot}\sim 10^{42}$ erg s$^{-1}$ \citep[e.g.,][]{Falcke2014}.

\section{KNBH binaries and FRB afterglows}

Recently, GW150914 was detected by the Laser Interferometer Gravitational wave Observatory (LIGO). The GWs were originated from the merger of a BH binary. The masses and spins of two initial BHs are $36^{+5}_{-4}~M_\odot$, $29^{+4}_{-4}~M_\odot$ and $<0.69\pm0.05$, $<0.88\pm0.10$, respectively, and the mass and spin of the BH after merger is $62^{+4}_{-4}~M_\odot$ and $0.67^{+0.05}_{-0.07}$, respectively \citep{Abbott2016,Zhang2016a}.

A binary system of BH might have a KNBH as one of its components (the younger one). When the holes are close to merge, the tidal forces should perturb the magnetosphere, which would then partially fall into the BH, neutralizing its charge and triggering a FRBs through the subsequent magnetic wave. Hence, a FRB might be a signal announcing an imminent GW burst. After the discharge of the BH, the field lines close to the rotation axis will reconnect sweeping away all residual plasma and ejecting a relativistic plasmoid. When such a plasmoid reaches the outer medium a shock will be formed. Such a shock can transform a part of the kinetic energy of the blob into internal energy in the form of relativistic particles, which might in turn cool through synchrotron and inverse Compton losses producing both radio and gamma-ray emission as in the external shock model of GRBs \citep[e.g.,][]{Gao2013}. These two steps are schematically represented in Figures 4 (a) and (b), which are similar to Figure 14 in \citet{Lehner2012}.

Recently, \citet{Punsly2016} proposed that the electric discharge of a meta-stable KNBH intermediate state would allow to operate the magnetic field shedding model of FRBs. In such a model the collapse of a magnetosphere onto a BH can generate a strong outward Poynting flux \citep{Hanami1997}, which should produce a radio and/or gamma-ray pulse.

In this scenario, the detectability of the FRB afterglow depends on the direction of the BH angular momentum (i.e., the rotation axis or the magnetic axis) and the ejecta opening angle. If the rotation axes of both BHs are almost aligned to the observer line of sight before the merger, a FRB and the subsequent afterglow might be detectable.

\begin{figure}
\centering
\includegraphics[angle=0,scale=0.4]{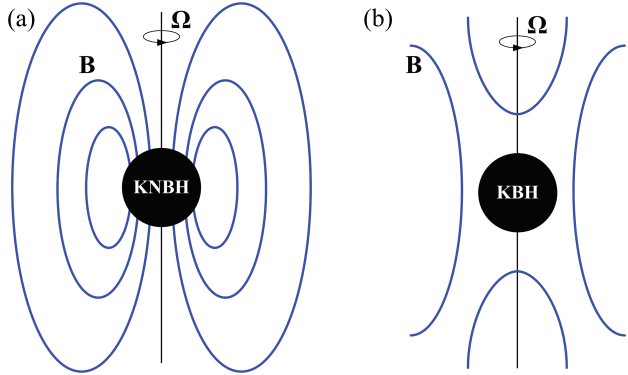}
\caption{Schematic diagram of the magnetic field configuration during the discharging of a KNBH in a binary.}
\label{fig4}
\end{figure}

\section{Discussion}

We suggest that the magnetospheric instability of a lone KNBH and a KNBH binary may result in FRBs and their afterglows.

In general, there are two possible ways of creating KNBHs. An isolated uncharged BH may be charged when it strays in the plasma environment, or a charged BH with oppositely charged magnetosphere may be the direct result of the gravitational collapse of a magnetized star \citep{Punsly1998}. The sudden discharge of these BHs through the instability of their magnetosphere should produce a FRB, but only in the case the BH spin is pointing nearly the observer a high energy counterpart should be observed. In addition, other mechanisms such as the implosion of a NS or a jet interaction with a turbulent low density plasma might also generate a similar phenomenology, at least in the radio domain.  We consider that the event rate of KNBH-induced FRBs should be only a fraction of total the event rate of FRBs, which is estimated to be around $10^{-3}~ \rm gal^{-1} ~yr^{-1}$ \citep[e.g.,][]{Thornton2013,Zhang2014}.

How can we differentiate between the mechanism here proposed and their competitors? The gravitational signal of colliding BHs in a binary might be a new multi-messenger channel to archive this. The \emph{Fermi} Gamma-ray Burst Monitor (GBM) recorded a weak gamma-ray transient 0.4 s after GW 150914 \citep{Connaughton2016}. Several models have been proposed to explain the possible electromagnetic counterpart of GW 150914 \citep[e.g.,][]{Li2016,Zhang2016a,Loeb2016,Perna2016}. As can be concluded from the above discussion, an alternative not invoking accretion might be related to the presence of a KNBH in the system. In such a case, a precursor FRBs might be detectable. The coordination of radio, gamma, and GW observations might result in a tool adequate to put to the test the ideas presented here: if a FRB is observed preceding a merger BH and it is followed by a short transient of high-energy radiation, we might rule out other possibilities such as direct NS collapse and coherent emission excited in ambient plasmas by a relativistic jet. In such a situation the present model should be strongly favored.

\section{Conclusions}

We proposed that charged and rotating BHs might be responsible for at least some FRBs when they discharge as a consequence of perturbations in their charged magnetospheres. Our model predicts that, if the right ambient conditions are present, the FRB might be followed by high-energy transients and a longer radio afterglow, similar to GRBs \citep[e.g.,][]{Liu2015a,Liu2015b,Luo2013,Hou2014,Song2015,Song2016}. In the case of BH binaries, if one of the holes is a KNBH surrounded by a magnetosphere, the FRB can be associated with a burst of GWs as the one recently detected by the LIGO and VIRGO Collaborations.
\\
\\
\\
\\
\emph{A Note Added.} Two days after this paper was posted in arXiv, Keane and his collaborators declared that they discovered FRB 150418 and a subsequent fading radio transient lasting $\sim$ 6 days \citep{Keane2016}. The transient can be used to identify the host galaxy. They concluded that the 6-day transient is largely consistent with a short GRB radio afterglow, but both its existence and timescale do not support progenitor models such as giant pulses from pulsars, and SNe. \citet{Vedantham2016} conducted the radio and optical follow-up observations of the afterglow, and argued that it may be associated with an AGN, not with FRB 150418, which is also discussed in literatures \citep[e.g.,][]{LiY2016,Williams2016}.

The isotropic energy of the afterglow is about $10^{50}$ erg and the beaming-corrected energy is below $10^{49}$ erg \citep{Zhang2016b}, which can be explained by synchrotron radiation as well as the external shock model in GRBs if the afterglow is associated with FRB 150418. Our model, on the other hand, can explain this event without invoking a GRBs nor an AGN.

\begin{acknowledgements}
We thank Bing Zhang, Brian Punsly, Wei-Min Gu, and Cui-Ying Song for helpful discussions, and the anonymous referee for very useful suggestions and comments. This work is supported by the National Basic Research Program of China (973 Program) under grant 2014CB845800, the National Natural Science Foundation of China under grants 11473022, 11475143, U1331101, and U1431107, and Science and Technology Innovation Talents in Universities of Henan Province under grant 14HASTIT043. GER is supported by grant AYA 2013-47447-C3-1-P (Spain).
\end{acknowledgements}

\end{document}